\documentclass[%
aip,
 amsmath,amssymb,
 preprint,%
 reprint,%
]{revtex4-1}

\usepackage{graphicx}
\usepackage{dcolumn}
\usepackage{bm}
\usepackage{siunitx}
\usepackage{xcolor}
\usepackage{booktabs}
\usepackage{float}

\usepackage[utf8]{inputenc}
\usepackage[T1]{fontenc}
\usepackage{mathptmx}
\usepackage{lineno}
\usepackage{soul}
\usepackage{natbib}

\begin{document}

\title[Enabling Integrated AI Control on DIII-D]{Enabling Integrated AI Control on DIII-D: A Control System Design with State-of-the-art Experiments}

\author{A. Rothstein}\email{arothstein@princeton.edu}
\affiliation{The two authors contributed equally to this paper}
\affiliation{Princeton University, Princeton, NJ, USA}
\author{H.J. Farre-Kaga}
\affiliation{The two authors contributed equally to this paper}
\affiliation{Princeton University, Princeton, NJ, USA}
\affiliation{Princeton Plasma Physics Laboratory, Princeton, NJ, USA}
\author{J. Butt}
\affiliation{Princeton University, Princeton, NJ, USA}
\author{R. Shousha}
\affiliation{Princeton Plasma Physics Laboratory, Princeton, NJ, USA}
\author{K. Erickson}
\affiliation{Princeton Plasma Physics Laboratory, Princeton, NJ, USA}
\author{T. Wakatsuki}
\affiliation{National Institutes for Quantum and Radiological Science and Technology, Naka, Japan}
\author{A. Jalalvand}
\affiliation{Princeton University, Princeton, NJ, USA}
\author{P. Steiner}
\affiliation{Princeton University, Princeton, NJ, USA}
\author{S.K. Kim}
\affiliation{Princeton Plasma Physics Laboratory, Princeton, NJ, USA}
\author{E. Kolemen}\email{ekolemen@pppl.gov}
\affiliation{Princeton University, Princeton, NJ, USA}
\affiliation{Princeton Plasma Physics Laboratory, Princeton, NJ, USA}

\begin{abstract}

We present the design and application of a general algorithm for Prediction And Control using MAchiNe learning (\texttt{PACMAN}) in DIII-D. Machine learing (ML)-based predictors and controllers have shown great promise in achieving regimes in which traditional controllers fail, such as tearing mode free scenarios, ELM-free scenarios and stable advanced tokamak conditions. The architecture presented here was deployed on DIII-D to facilitate the end-to-end implementation of advanced control experiments, from diagnostic processing to final actuation commands. This paper describes the detailed design of the algorithm and explains the motivation behind each design point. We also describe several successful ML control experiments in DIII-D using this algorithm, including a reinforcement learning controller targeting advanced non-inductive plasmas, a wide-pedestal quiescent H-mode ELM predictor, an Alfv\'en Eigenmode controller, a Model Predictive Control plasma profile controller and a state-machine Tearing Mode predictor-controller. There is also discussion on guiding principles for real-time machine learning controller design and implementation. 

\end{abstract}
\maketitle

\section{Introduction}\label{sec:intro}

Magnetic confinement fusion has the potential to provide sustainable, carbon-free energy to meet an increasing global energy demand. Among candidate reactor designs, the tokamak is considered one of the most promising for near-term commercialization \cite{buttery_advanced_2021}. Because tokamak plasmas are inherently dynamic, real-time feedback control is essential for achieving and maintaining advanced operational regimes  in devices such as ITER\cite{hassanein_potential_2021}, DEMO\cite{zohm_assessment_2013}, and SPARC\cite{creely_overview_2020}.The design of real-time plasma control systems (PCS) has therefore been the subject of sustained development across multiple devices \cite{lennholm_plasma_2000,kudlacek_overview_2024,yan_custom_2021,yonekawa_current_2004,margo_current_2020,galperti_overview_2024,de_vries_strategy_2024,perek_preliminary_2025,creely_overview_2020,creely_sparc_2023,hahn_advances_2020}. 

Plasma instabilities evolve on a wide range of timescales, from the millisecond order for vertical displacement events (VDEs) \cite{hassanein_vertical_2008} to over a hundred milliseconds for profile control. To address these requirements, modern PCS architectures employ multiple CPU cores operating at different cycle times, each dedicated to specific control objectives\cite{margo_current_2020,hahn_advances_2020}. A key component of nearly all PCS systems is a real-time equilibrium reconstruction which provides the plasma shape and boundary information needed for fast position and shape control\cite{ferron_real_1998,moret_tokamak_2015}. 

The field of plasma physics has seen a surge in applications of machine learning (ML) across a wide range of tasks\cite{anirudh_2022_2023} including ML for real-time control. One common approach is to develop surrogate models of computationally expensive physics codes\cite{morosohk_accelerated_2021,morosohk_neural_2021,morosohk_machine_2024,rothstein_initial_2024,rothstein_torbeamnn_2025,shousha_machine_2023,kim_highest_2024,gupta_detachment_2025} which is then integrated with conventional controllers. Beyond surrogates, ML methods have been applied to instability suppression and avoidance and profile prediction and control\cite{jalalvand_real-time_2021,abbate_data-driven_2021,montes_machine_2019,rea_real-time_2019,guo_disruption_2021,fu_machine_2020,hu_real-time_2021,morosohk_estimation_2022,morosohk_realtime_2023,chang_dltm_2025}. These event prediction methods are typically coupled with basic event thresholding or finite state controllers. Finally, the current state-of-the-art ML controllers utilize reinforcement learning (RL) trained in offline simulator environments before deploying in a test reactor environment\cite{degrave_magnetic_2022,seo_avoiding_2024,wakatsuki_safety_2019,char_offline_2023,de_tommasi_rl-based_2022,dubbioso_deep_2023}. 

Despite these advances, most ML-based controllers have been implemented as stand-alone demonstrations. Little effort has been devoted to creating a common, real-time compatible framework that simplifies deployment across multiple and diverse control applications. This presents a barrier to scaling ML-based control beyond one-off projects.

\begin{figure*}
    \centering
    \includegraphics[width=0.9\textwidth]{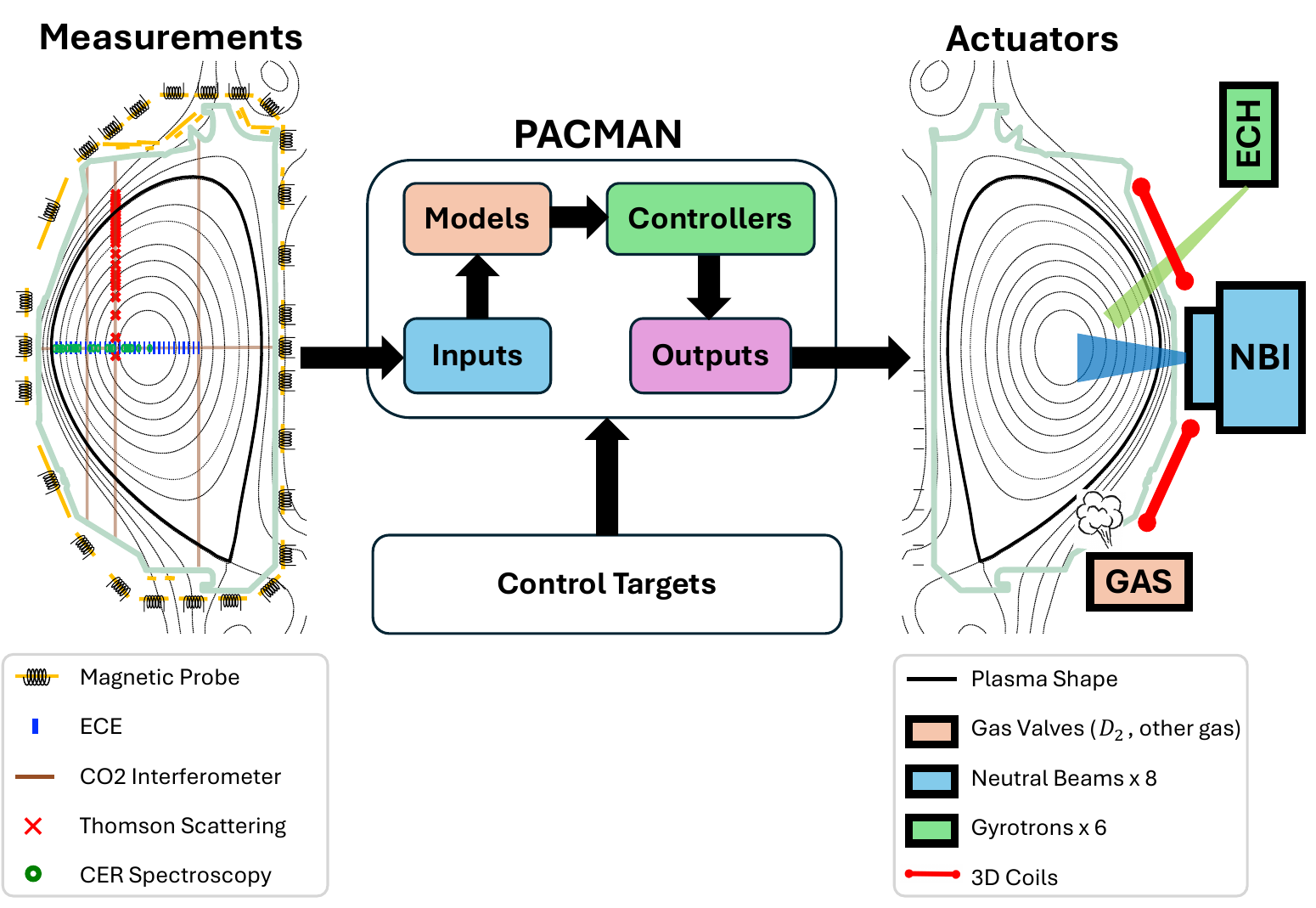}
    \caption{DIII-D diagnostics, actuators and PACMAN. PACMAN receives diagnostic data and finds the optimal actuators to accomplish the user defined control targets. Location of diagnostics and actuators are approximate in this sketch. }
    \label{fig:diii-d_PACMAN}
\end{figure*}

This work introduces such a framework, shown in Figure \ref{fig:diii-d_PACMAN}, called \texttt{PACMAN} (Prediction And Control using MAchiNe learning), and reports its first implementation on DIII-D. \texttt{PACMAN} builds on lessons from a prior profile-control oriented workflow \cite{abbate_general_2023}, addressing its limitations in scalability, clarity of model-controller separation, diagnostic integration (new diagnostic signals), and timebase management. The design emphasizes modularity, fault-tolerance, and flexibility, making it possible to accommodate a wide range of ML models and controllers. Although demonstrated in a CPU-based environment, the principles are sufficiently general to be extended to PCS workflows that include other hardware such as GPUs \cite{boyer_toward_2021,rath_high-speed_2012,rath_adaptive_2013,huang_implementation_2016} and FPGAs\cite{wei_low_2024,le_new_2014}.

The remainder of this paper is organized as follows: Section \ref{sec:algorithm} presents an overview of the \texttt{PACMAN} architecture and its design rationale. Section \ref{sec:exp} describes five experimental applications implemented within this \texttt{PACMAN} framework on DIII-D. Finally,  Section \ref{sec:conclusion} summarizes the contribution, its potential limitations and outlines future directions. 

\section{Algorithm Design}\label{sec:algorithm}

\begin{figure*}
    \centering
    \includegraphics[width=\textwidth]{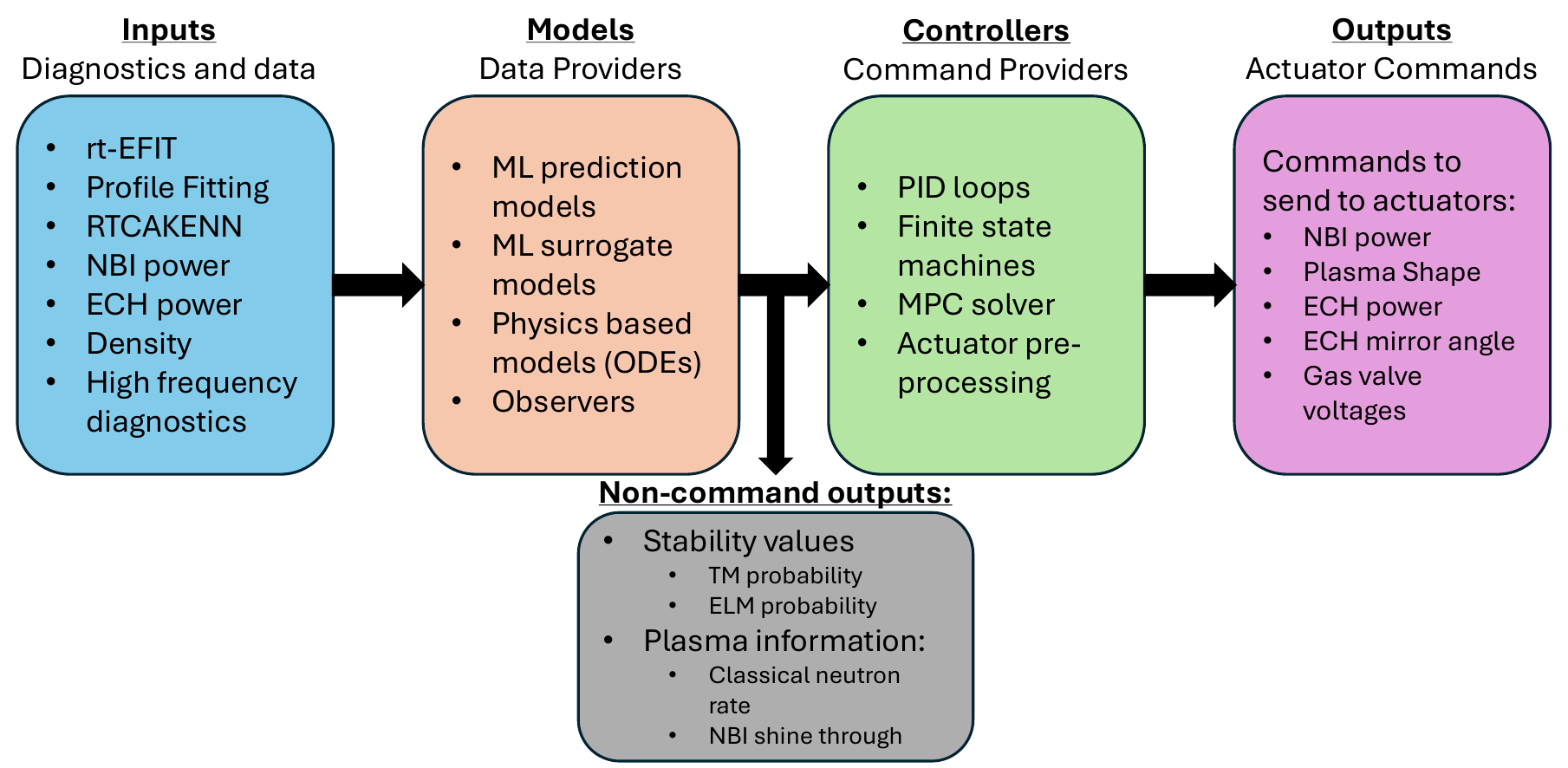}
    \caption{General design of the \texttt{PACMAN} algorithm. The algorithm is a end-to-end controller, starting from diagnostic inputs, followed by ML, standard control, or physics-based models, then saving the ML outputs and sending the data to a controller such as finite state machines and PIDs, and ending with commands to actuators such as neutral beam injector (NBI) power, electron cyclotron heating (ECH) power, and gas requests. }
    \label{fig:general_design}
\end{figure*}

The \texttt{PACMAN} algorithm is outlined in 
Figure \ref{fig:general_design} and consists of four blocks: 
\begin{enumerate}
    \item Diagnostic input pre-processing
    \item Model block
    \item Controller block
    \item Actuation output post-processing
\end{enumerate}

While the subsequent subsections (Section \ref{sec:inputs}, \ref{sec:model},\ref{sec:controller},\ref{sec:outputs}) will describe each block in detail, here we will explain the high level design decisions that went into the overall architecture.   

The first key design principle is the separation of each block and the flow of data and information forward from one block to the next. The models can utilize any of the inputs, the controllers can utilize any data from the models and inputs, and finally the actuator block can utilize all preceding data. A key separation is that components within a block cannot communicate within the block. For example, models cannot communicate with other models and similarly controllers cannot communicate with other controllers. If a model needs the output of another model, these would be jointly considered a single model, with the same approach also applied to integrated controllers that need to communicate. Multiple controllers that have to interact would be considered a single ``Controller". This important requirement was made to avoid possible dependency collisions where some models must complete running before others can start. This enables a future option of running all models in parallel followed by all controllers in parallel, greatly speeding up the entire real-time pipeline. Error checking is centralized at block interfaces, reducing overhead and clarifying where faults are detected and handled.

\begin{figure*}
    \centering
    \includegraphics[width=\linewidth]{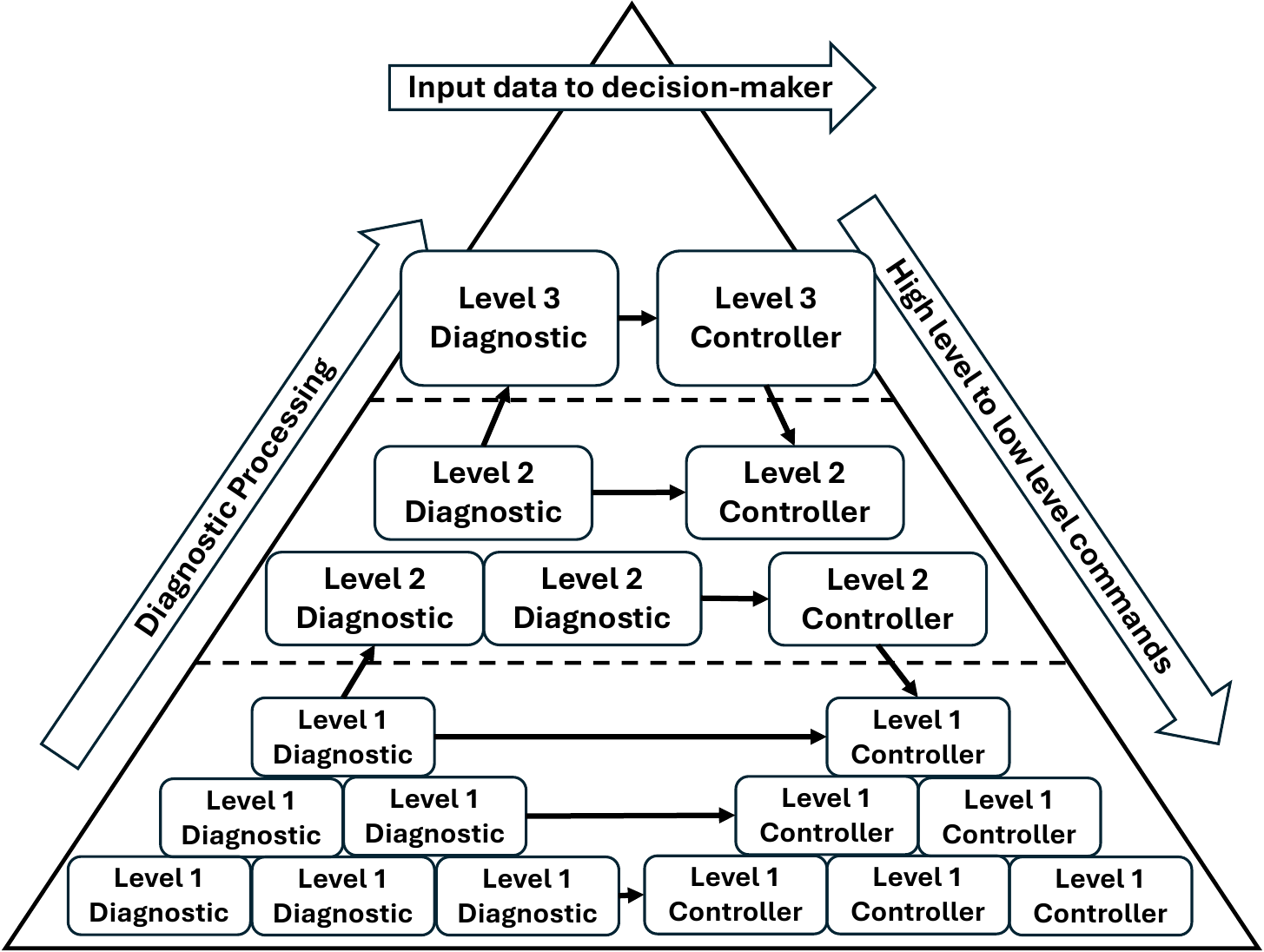}
    \caption{3 level design structure of the PCS diagnostics and algorithms. The flow of diagnostic information is always up levels and to the right, where raw diagnostic data is processed, then may be fed to automated fitting routines such as real-time EFIT\cite{ferron_real_1998}, before being provided as inputs to various controllers. On the controller side, commands travel down the levels from the highest controllers, which may decide on target profile objectives, or plasma stability targets, and gets passed down to lower level controllers that will achieve those high level goals before the lowest level controllers calculate exact commands to be send to hardware. Of note, as seen in the figure there are typically more Level 1 diagnostics and controllers  than there are Level 2 or 3. \texttt{PACMAN} is at the top of the pyramid as a Level 3 controller. }
    \label{fig:levels}
\end{figure*}

The next key principle is the idea of a 3 level algorithm design. In this paradigm, algorithms are classified into 3 levels with distinct purposes and communicate with algorithms in adjacent levels. These levels are defined as:
\begin{enumerate}
    \item \textbf{Level 1}: Lowest-level algorithms to calculate hardware commands from more generic actuator commands or directly process diagnostic data. Examples here include calculating exact power supply voltage requests from desired coil currents or reading digitizer voltages. A typical user will never interact with a Level 1 algorithm. 
    \item \textbf{Level 2}: Mid-level algorithms that process raw diagnostic data or turn high level actuator commands into lower level commands. Examples here include fitting Thomson scattering data to produce $n_e$ profiles or computing On/Off modulations of NBIs. A typical user will have some interactions with a Level 2 algorithm. 
    \item \textbf{Level 3}: High-level algorithms that process data and compute high level actuator commands to achieve a desired control objective. Examples here include most standard control such as PID-based $\beta_N$ control as well as more advanced controllers. A typical user should have the majority of their interactions with Level 3 algorithms. 
\end{enumerate}

This design closely mimics the human interaction in a typical tokamak control room.  Teams of engineers own the actuators controlled by Level 1 algorithms. Physics operators, akin to a Level 2 algorithm, bridge the gap between experimentalists and the technical staff that operate the machine. Experimentalists nominally existing at Level 3 ask physics operators to conduct experiments on their behalf. This separation of concern not only provides a metaphorical comparison between software and users but mimics the possible administrative constraints in the control room that allow effective tokamak operation. For example, a physics operator might tell an experimentalist that a particular plasma scenario is out of reach, and similarly a Level 2 algorithm may accept or reject a total beam power request based on configured limits.

While this is the design standard is not enforced as the standard for algorithm design on DIII-D, it was first introduced on the NSTX PCS and has been attempted to be used as a general design principle. Because of this, it is worth noting that this design principle does not hold across all algorithms in the DIII-D PCS and there are edge cases of algorithms crossing into 2 levels. \texttt{PACMAN} sits cleanly in Level 3 and takes control objectives from the user combined with processed data provided by Level 2 algorithms and calculates actuator commands that are sent down to Level 2 algorithms. 

Moving towards discussion on each individual block, their individual design decisions were made to achieve the following major objectives:
\begin{itemize}
    \item Scalable to many models and controllers where all can be run concurrently
    \item Agnostic to type of ML model or controller type
    \item Standard processing of inputs that is still flexible
    \item Rough cycle time goal of \SI{5}-\SI{50}{ms} range. All should run in this range even if they take performance hits
    \item Actuator arbitration and handling of conflicting actuator commands
\end{itemize}

\subsection{Input Gathering and Pre-processing}\label{sec:inputs}

\begin{table*}[]
    \centering
    \begin{tabular}{|c|c|c|c|}
        \hline
        \textbf{Input} & \textbf{Source} & \textbf{Time Scale (ms)} & \textbf{Pre-processing} \\\hline
        Plasma Shape Information & Real-time EFIT\cite{ferron_real_1998} & <1 & \texttt{RTEFIT} \\
        $B_T$ & Rogowski coils & <1 & $B_T=k\times I_{E coil}$ \\
        $I_P$ & Rogowski coils & <1 & \\
        NBI power and torque & NBI Algorithm\cite{pawley_advanced_2017} & 10 & LP Filter\\
        ECH power & ECH Algorithm\cite{cengher_status_2020} & 10 & \\
        ECH mirror angles & ECH mirror encoders\cite{kolemen_real-time_2013} & 20 & \\
        ECH and ECCD deposition & Real-time TORBEAM\cite{poli_torbeam_2018} & 20 & \\
        Real-time Kinetic Profiles & \texttt{RTCAKENN}\cite{shousha_machine_2023} & 5 & \\
        Average $n_e$ & CO$_2$ interferometer & 1 & \\
        Current profile measurements & Motional Stark Effect\cite{holcomb_overview_2008} & 10 & \\
        Ion temperature, density, and rotation profiles & Charge Exchange Recombination\cite{chrystal_improved_2016} & 20 & \texttt{Profile Fitting} \\
        Electron temperature and density profiles & Thomson Scattering\cite{ponce-marquez_thomson_2010} & 20 & \texttt{Profile Fitting} \\
        Local $n_e$ fluctuations & Electron Cyclotron Emission\cite{austin_electron_2003} & 0.001 & \\
        Global $n_e$ fluctuations & CO$_2$ Interferometer Fluctuations & 0.001 & \\
        Turbulence Measurement & Beam Emission Spectroscopy & 0.001 & \\\hline
        \textbf{Outputs} & \textbf{Source} & \textbf{Time Scale (ms)} & \textbf{Post-processing} \\\hline
        NBI Power & \texttt{VEP}\cite{boyer_feedback_2019} & 10 & LP Filter \\
        NBI Torque & \texttt{VEP}\cite{boyer_feedback_2019} & 10 & LP Filter \\
        Individual NBI dutycycles & \texttt{VEP}\cite{boyer_feedback_2019} & 10 & LP Filter \\
        ECH power & Gyrotron Algorithm\cite{cengher_status_2020} & 10 & LP Filter \\
        ECH mirror angle & Real-time TORBEAM\cite{poli_torbeam_2018,kolemen_real-time_2013} & 100 & \\
        Gas voltages & Gas Algorithm & 1 & \\
        Shape targets & Proximity Control\cite{barr_development_2021} & 1 & \\
        RMP Current & 3D Field Algorithm\cite{evans_rmp_2008} & 1 & \\
        RMP Phasing & 3D Field Algorithm\cite{evans_rmp_2008} & 1 & \\
        RMP n number & 3D Field Algorithm\cite{evans_rmp_2008} & 1 & \\
        \hline
    \end{tabular}
    \caption{List of all inputs that can be used as model inputs within \texttt{PACMAN} and all possible output actuators. The timescales of inputs are how often new diagnostic data is received or processed by external algorithms, i.e. in the case of ion temperature profiles the \SI{20}{ms} timescale is how often \texttt{Profile Fitting} computes new fitted profiles, not how often new diagnostic data is produced. The actuator timescales are based on how quickly actuator effects can affect the plasma. While ECH power can be turned on in a sub-millisecond timescale, to achieve continuous modulated power requires waiting closer to \SI{10}{ms} to get a well time-averaged power. }
    \label{tab:in-out}
\end{table*}

The first block in \texttt{PACMAN} gathers and standardizes input data to further process it in the subsequent blocks. At the time of writing, the full list of real-time inputs at DIII-D is given in the top half of Table \ref{tab:in-out}. However, new diagnostics are constantly added to DIII-D while other currently offline-only diagnostics are made real-time capable as well. In order to use new diagnostics at different time scales, and data sizes, it is necessary to design the input block to be adaptable to new diagnostics. 

Diagnostics run on timescales varying from high frequency (\SI{1}{\MHz}) down to \SI{100}{\Hz}, or every \SI{10}{\ms}. Storing and buffering this quantity of data is intractable in a real-time system that is expected to function reliably. Note that this is not a problem for \texttt{PACMAN} specifically, but is an issue that any conventional real-time control system must deal with. We let \texttt{PACMAN} rely on the Level 2 data providers to handle the data buffering on their own CPUs and then read the latest buffer information. This streamlines the data collection process and improves \texttt{PACMAN}'s adaptability to new diagnostics by not needing to handle individual data buffering internally. \texttt{PACMAN} saves the most recent cycle of data, typically the previous \SI{10}{ms} or \SI{20}{ms}, to provide as reference for how the plasma state is evolving. \texttt{PACMAN} does not save data over a longer time horizon because that is expensive in terms of complicating the \texttt{PACMAN} algorithm and expensive in terms of computational cost. The responsibility to buffer data is on any level 2 diagnostic to buffer relevant data and provide it to \texttt{PACMAN}'s input block. 

General pre-processing is kept as minimal as possible to leave ML models to interact with the data closest to the diagnostics. One of the major signal pre-processing computed is the low pass filtering of signals such as NBI power and torque, $P_{NBI}$ and $T_{NBI}$, respectively, as well as ECH power, $P_{ECH}$, which achieve continuous power control by quickly alternating power on and off at fixed amounts. The toroidal field is approximated by taking the current from the B-coil ($I_{B_{coil}}$) and multiplying by a geometric factor for the DIII-D configuration of $B_T=\frac{144\mu_0}{2\pi \times \SI{1.6955}{\m}}I_{B_{coil}}$. We do not normalize any signals in a custom way, such as $z$ or (0, 1) normalization as it is required for ML models, because the exact method of custom normalization depends on the actual model. Thus, these kinds of data-dependent normalizations are part of the model block to maintain flexibility for every model to use the normalization methods that were found most suitable during the ML model development.

Additionally, the input block must check for real-time errors in all of the diagnostic signals and inform downstream models and controllers of diagnostic failures. These checks come in a variety of types and are specific to what is providing the data. In the case of real-time EFIT, there is a built-in error waveform to inform if the calculation converged. Other checks can involve checking if a temperature profile is greater than 0 or if the timestamp of the most recent diagnostic data is too old. These checks are important for two reasons: first they provide a decision pipeline for models to decide if they will run without correct input data (usually they will not), and second, this provides insight in the control room that specific input data had issues during the last shot and must be addressed before the next shot. PACMAN is robust to diagnostic errors, so a model/controller will simply not run if an essential diagnostic is unavailable for that timestep. 

In summary, the Input block in \texttt{PACMAN} provides error-checked real-time data that has been gathered from Level 2 algorithms as well as the data from the previous cycle. There is minimal pre-processing to maximize flexibility of the models in the next block.

\subsection{Model Block}\label{sec:model}

The next section in the \texttt{PACMAN} algorithm is the model block. This is where any typical ML model will cover tasks that involve taking data from the input block and producing values actionable by the controller block. To be clear, for the applications mentioned here we describe the use of ML models in this block, but this block can execute non-ML models that meet the requirements to function here. 

Every model block receives input/output information as well as custom parameters for normalization and standardization of the input data, information about the model such as the model type and the number of models. Every model block runs through the following steps:

\begin{enumerate}
    \item All information is passed to the model block via a structured data format that needs to be parsed task-dependently.
    \item The input data is preprocessed. This involves custom pre-processing steps like filtering/smoothing and normalization.
    \item The actual ML model is instantiated. Since PACMAN is designed model-agnostic, a wide range of implementations of ML models is supported. Several ML models and functions may be run in series within the model block. 
    \item All the errors produced from the input block are evaluated to make sure the input data is valid. Only if this is successful, the subsequent steps are executed in order to provide machine protection for the tokamak.
    \item The preprocessed input data is passed through the ML model and the raw outputs are collected.
    \item The raw outputs are customly post-processed. This can again involve smoothing/filtering, or also thresholding for binary tasks.
    \item The post-processed output is written back into the structured data format such that the subsequent controller block can parse it.
\end{enumerate}  

Within PACMAN, we support a large variety of ML models. Currently, this includes: ML surrogate models, event prediction models, plasma profile prediction models, observers, latent space mappings, or RL models. More models for new tasks are steadily integrated. It is important to note the flexibility of PACMAN in the sense that all models are responsible for handling their custom normalizations and any model-specific processing of inputs before running the specific model.

Currently, PACMAN supports multiple implementations for ML models in real-time such as \texttt{keras2c}\cite{conlin_keras2c_2021}, \texttt{PCAnet}\cite{morosohk_accelerated_2021}, reservoir computing networks (RCN)\cite{rothstein_initial_2024}, or more classical approaches with less ML such as linear regression models, random forests\cite{boyer_toward_2021}, or standard mathematical functions. Within this structure, we have leveraged the flexibility to implement a more complex deep survival machine architecture\cite{nagpal_auton-survival_nodate} to help with tokamak event prediction. 

While \texttt{keras2c} provides an easy path to PCS model implementation, it does so at significant cost to stack memory use and does not scale well to many \texttt{keras2c}-generated functions running concurrently. This is mainly due to developer time constraints and this problem could in theory be solved with additional focused code development. Additionally, this limits the total size of a model that is implementable, with the limit of a DIII-D \texttt{keras2c}-generated function file being approximately \SI{3}{MB}. The advantage of using \texttt{keras2c} is the ability to use more complex ML layers, such as convolutional layers and LSTMs, while other implementations that scale better by allocating memory on the heap rather than the stack, such as PCAnet, limit the user to just MLP architectures. 

As described previously, all models in the model block are fully independent and may be run in series or parallel depending on computing hardware availability. While this limits the ability for models to contribute to inputs for other models, this decision enables the future ability to provide a significant computation speed-up by enabling parallelization. If a user desires some type of ML hierarchical model that wants to use outputs from multiple models, such a model could be implemented as multiple \texttt{keras2c}-generated functions within the same single Model sub-block. For example, the survival models mentioned are comprised of 4 \texttt{keras2c}-generated functions run as one Model.  

The model block provides a variety of outputs to be used in the controller block or other parts of the PCS. The simplest are computed values needed by other algorithms external to \texttt{PACMAN}. Examples here could be surrogate models that calculate physics values such as NBI shine-through, MHD stability metrics, ECH ray trajectories, or plasma profile estimates. Those model outputs can also go directly to the controller block for basic control processing such as PID controllers, thresholding-based decision making, or finite state machines. Other model block outputs can include outputs that require additional processing or advanced input processing for the specific controller such as an advanced plasma state mapping used in Section \ref{sec:MPC}.

\subsection{Controller Block}\label{sec:controller}

The controller block can be considered the `decision maker' of the \texttt{PACMAN} algorithm. Each module in the controller block will receive the necessary outputs from the model block, as well as any required input block data, and calculate actuator commands to send to the output block. Again following the structure of the model block, none of the controllers can interact with any other controller within the controller block and any integrated controller would be a single ``Controller" with multiple internal algorithms running. This enables future versions of \texttt{PACMAN} to parallelize controllers because they can all run independently. 

This restriction on separation controllers may limit possible future integrated controllers, but have decided the limitations are worth the safety provided in avoiding possible conflicting actuator commands. Future applications that aim for more integrated control  within \texttt{PACMAN} could turn on and off specific controllers based on what part of the shot is occurring. The user may utilize specific controllers in $I_P$ flattop while others may be used during ramp-up and ramp-down sequences. 

Example controllers can be finite state machines\cite{shousha_design_2022}, basic thresholding decision making, PID control, MPC optimizers, other advanced control algorithms, or anything that will produce an actuator request from model outputs or diagnostic data. Each module in the controller block can provide any subset of possible actuators listed in Table \ref{tab:in-out}, and multiple controllers can attempt to control the same actuators. Collisions between actuator commands are handled within the outputs block. 

A final note is the controllers are aware of previous real-time errors from inputs as well as models. If any of the inputs for a specific controller have an error, whether from the model block or input block, then the controller will not compute any actuator command. This is done to maintain machine safety in an abundance of caution, but of note sometimes inaction can potentially be more dangerous than a bad action. In the future, other default actions could be created such as designated ramp-down sequences to safely shutdown the plasma. 

\subsection{Actuator Output and Post-Processing}\label{sec:outputs}
Finally the output block gathers commands from the controller block, imposes additional machine safety constraints, and sends commands to Level 2 algorithms to handle actuation. The list of \texttt{PACMAN} controllable real-time  actuators on DIII-D, at present, is listed in the bottom half of Table \ref{tab:in-out}. Similar to the table of inputs, \texttt{PACMAN} has been designed to handle future new actuators being added in anticipation of advancements on DIII-D, for example shattered pellet injection. While some actuators can change quicker than \texttt{PACMAN}'s millisecond evaluation timescale, \texttt{PACMAN} cannot send commands faster than it can run. This limits the speed at which new actuation commands can be given, but at the millisecond timescale that is sufficient for most plasma instabilities of interest. \texttt{PACMAN} would be inappropriate for controllers targeting events like VDEs which can evolve and disrupt on sub-millisecond timescales. 

First, the output block checks for any actuator collisions from the different controllers. If the output block receives two conflicting actuator commands, no actuation is output to maintain machine safety, and the operators are informed of the controller actuation conflict. While this approach is naive at present, the separation and flexibility provided by the output block enable future improvements to actuator conflicts. Multiple controller actuation can be added together where one model provide a `base load' for an actuator like $P_\mathrm{NBI}$ and another model provides a minor correction with some $\Delta P_\mathrm{NBI}$. 

Finally before any commands are sent, final clamps on actuator commands are done to set the minimum and maximum values of the commands. While many of these safety limits exist at various lower levels the DIII-D PCS hierarchy, we include them here to guarantee that no bad actuator commands will be executed. This includes setting minimums and maximums for gas valve voltages, $P_{NBI}$, $P_{ECH}$, and limits to ECH mirror angles and 3D coil currents. 

\section{Experimental Applications}\label{sec:exp}
\begin{figure*}
    \centering
    \includegraphics[width=\linewidth]{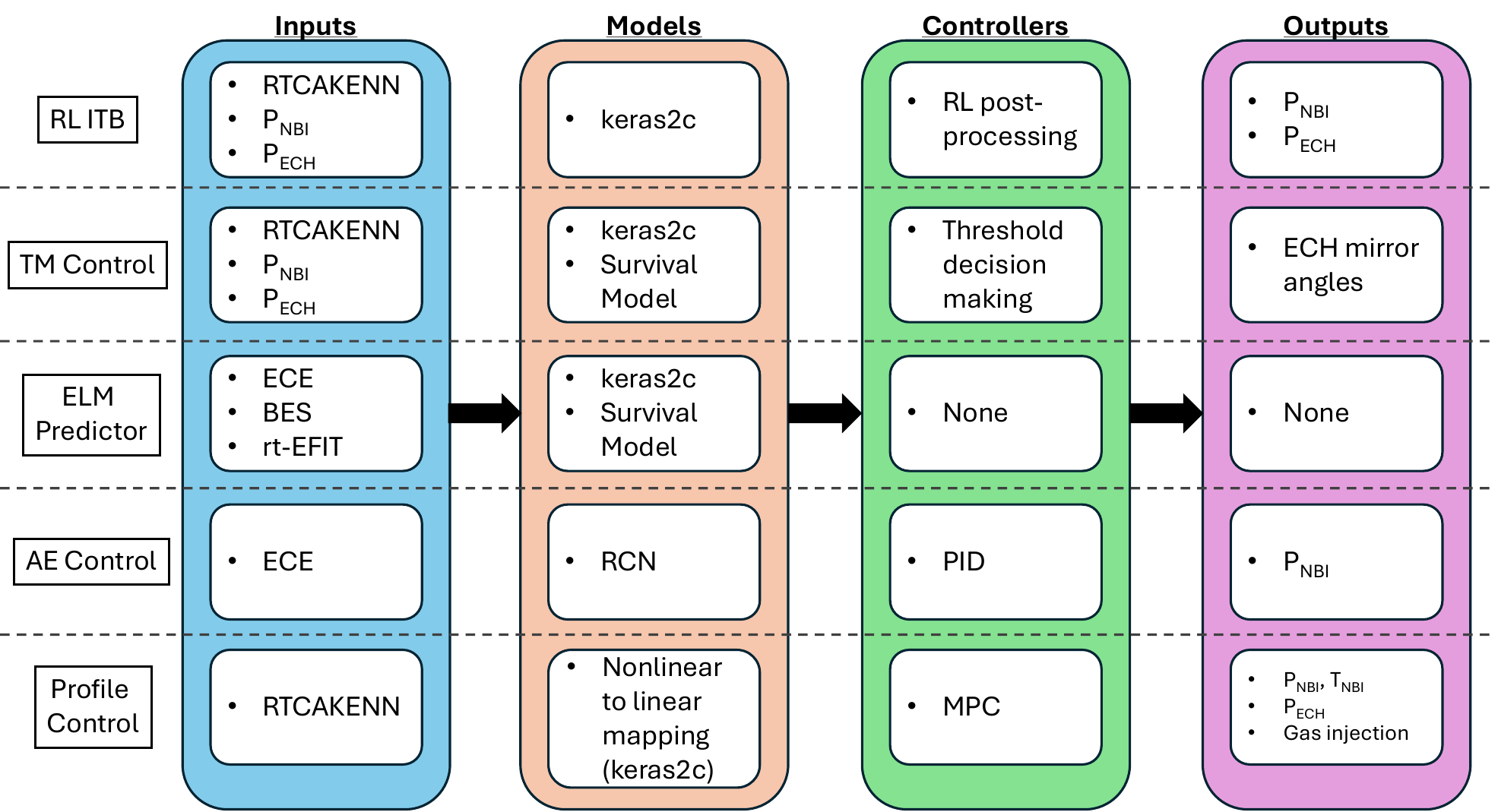}
    \caption{Overview of the 5 experimental applications that are described. Each block is depicted mirroring the layout from Figure \ref{fig:general_design}, where each horizontal dashed line separates each experiment. Note the many shared inputs and outputs that utilize the framework. }
    \label{fig:experiments}
\end{figure*}

\begin{table*}[]
    \centering
    \begin{tabular}{c|c|c|c}
        \textbf{Model/Controller Type} & \textbf{Objective} & \textbf{Actuator} & \textbf{Inference Time (ms)} \\\hline
        RL Controller & $\beta_N$ and ITB Control & $P_{NBI}$ and $P_{ECH}$ & 
        0.8 \\
        TM Controller & TM Avoidance & ECH mirror angle & 0.6 \\
        ELM Predictor & ELM Prediction & None & 0.2\\
        AE Controller & Suppress AEs & $P_{NBI}$ & 0.65 \\ 
        Latent-linear MPC & Profile Control & $P_{NBI},T_{NBI},P_{ECH}$ and gas injection &  
        15
    \end{tabular}
    \caption{Overview of the 5 \texttt{PACMAN} experimental applications described in Section \ref{sec:exp}. Each control type is listed along with the control objective, the actuators used, and the time scale of the ML model execution.}
    \label{tab:exp-overview}
\end{table*}

In this section we will highlight some of the successful applications of the \texttt{PACMAN} algorithm for real-time plasma control. These applications are summarized in Table \ref{tab:exp-overview}. Each algorithm fits neatly into the block described in Section \ref{sec:algorithm} by the visualization in Figure \ref{fig:experiments}. While there are additional algorithms that have been deployed in experiment on the DIII-D PCS using the \texttt{PACMAN} framework, the examples given here highlight the flexibility of the architecture to enable deployment of various different ML model and controller types to achieve a diverse set of control objectives. 

Each experiment described in this section will feature: a  description of the model or controller objectives, the list of model inputs and outputs, and finally discuss what computations occur in the model and controller blocks, respectively.

\subsection{Reinforcement Learning Controller}\label{sec:RL}

\begin{figure*}
    \centering
    \includegraphics[width=\linewidth]{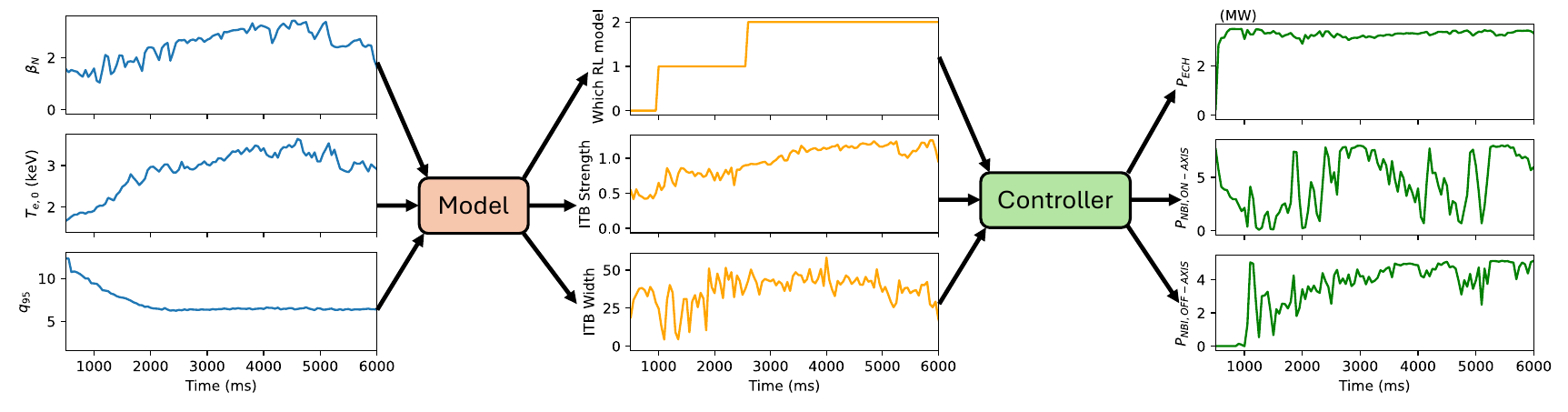}
    \caption{DIII-D shot 204975 where the RL ITB controller was used to control $\beta_N$ and ITBs. On the left are three of the real-time inputs: $\beta_N$, core electron temperature, and the $q$ profile at $\psi_N=0.95$. These along with the remaining inputs are fed into the model where the mdoel outputs information on which RL model was selected as well as ITB strength and width. This information is fed into the controller block that finalizes the ECH, on-axis NBI power, and off-axis NBI power that get sent to the actuators. }
    \label{fig:RL_ITB}
\end{figure*}

The first example controller is an RL based $\beta_N$ and internal transport barrier (ITB) controller which demonstrates `full ML control' where the outputs of an ML algorithm fully decide actuator commands, subject to hard coded machine safety actuator limits. This was originally trained for the DEMO reactor\cite{wakatsuki_safety_2019} and a similar RL controller was trained offline using the RAPTOR code\cite{felici_non-linear_2012} as a simulator environment. The control objectives were: follow a $\beta_N$ target trajectory and excite ITBs in flattop in the high q$_\textrm{min}$ advanced tokamak scenario. After offline training was completed, the hierarchical RL model was converted to 4 \texttt{C} functions using \texttt{keras2c}. For more details on the controller and experimental results, look to \citet{wakatsuki_RL}. Real-time time traces of some inputs, some model outputs, and final actuator commands can be seen in Figure \ref{fig:RL_ITB}. 
 
The RL model was trained to take in the q and $T_e$ profiles along with present actuator information. There was additional input pre-processing added to compute the magnetic shear, $s2=2\frac{\psi_N}q\frac{\mathrm dq}{\mathrm d\psi_N}$, from the q profile and to calculate on and off axis NBI powers from individual NBI powers. The model block outputs request for $P_\textrm{NBI,on-axis}$, $P_\textrm{NBI,off-axis}$, and $P_{ECH}$. For the control objectives, this RL controller needed \texttt{PACMAN} to run at \SI{50}{ms} cycle time and this was achieved.

The \texttt{keras2c}-generated functions were placed into the model block of \texttt{PACMAN} and given the appropriate inputs from the input block. The hierarchical structure used a first model to observe information about the ITBs and to produce an index to be used by the controller to decide which RL model function to use. Using this index, one of three  \texttt{keras2c}-generated functions was selected to run. This secondary model took the same inputs as the observer and produced output signals of $P_\textrm{NBI,on-axis}$, $P_\textrm{NBI,off-axis}$, and $P_{ECH}$ to be passed to the controller block for further processing. 

The controller block for the RL ITB model took the outputs from the final \texttt{keras2c}-generated function and converted them to appropriate commands for the Level 2 algorithm in the PCS. The NBI control algorithm in the DIII-D PCS cannot receive targets for $P_\textrm{NBI,on-axis}$ and $P_\textrm{NBI,off-axis}$ directly, so the controller block was used to convert $P_\textrm{NBI,on-axis}$ and $P_\textrm{NBI,off-axis}$ to appropriate individual NBI powers to achieve those actuator target commands. 

\subsection{TM control with ECCD}\label{sec:TM}

The next controller utilized a real-time tearing mode (TM) predictor model to steer the ECH mirror and preemptively suppress TMs. This model was trained using the Auton Survival Model  framework\cite{nagpal_auton-survival_nodate} and the controller was tested experimentally in the elevated q$_\textrm{min}$ advanced tokamak scenario. For a complete detailing of the ML model and control results, look to \citet{rothstein_preemptive} and \citet{farre_interpreting}. 

The inputs for the TM predictor model were the profiles provided by \texttt{RTCAKENN}, $P_{NBI}$, $P_{ECH}$, and various scalar parameters from rt-EFIT. The output of the controller block is an ECH mirror objective: a $\rho$ aiming target for broad off-axis current drive or the $q=2$ rational surface for TM suppression. With \texttt{PACMAN} sending one of these ECH mirror objectives, the Level 2 algorithms of the real-time TORBEAM combined with real-time ECH mirror steering can find the optimal ECH mirror angle to achieve the ECH mirror target. Due to slow actuation timescale of ECH mirror steering, the target inference time for this controller was under \SI{10}{ms} and easily achieved by \texttt{PACMAN}.

The model block contains the \texttt{C} implementation of the Deep Survival Machine in the Auton Survival package with \texttt{keras2c}-generated functions. The internal math functions used were made real-time compatible and avoid any possible overflow errors from standard ML functions like \texttt{LogSoftMax} and \texttt{LogSoftExp}. The final output of the model block is a TM probability over some time horizon, where the time horizon is a configurable user input. 

Finally, the controller block of the TM controller uses a basic thresholding scheme to select gyrotron tasks, also refered to as a finite state machine. For example, when the TM probability rises above 10\%, the first gyrotron will switch from off-axis current drive to $q=2$ surface tracking. If the TM probability rises above 20\%, then the second gyrotron will switch tasks and so on. There is also an additional low pass filter on the TM probability to smooth the signal and avoid changing the actuator significantly based on one possible poor data in a specific CPU cycle. Combining this control scheme with an ML predictor model enables a control scheme based on physics intuition of TM control, while leveraging advanced ML techniques to predict stability in real-time. 

\subsection{ELM Prediction}\label{sec:ELM}
The third application example of this framework was in sporadic breakthrough ELM prediction in wide-pedestal QH mode (WPQH). Unlike the other applications, this example utilized the \textit{non-command outputs} path to leverage \texttt{PACMAN}'s out-of-the-box ready processing of real-time diagnostic data to plug into an already developed external controller. This model and controller were deployed and tested in an experiment for WPQH pedestal control, the details of which are presented in an upcoming paper by \citet{butt_elm}. 

Sporadic breakthrough ELMs in intrinsically ELM-free plasmas is a phenomenon not yet well understood from a physics perspective, so a simple predictive model does not yet exist. Since over a decade of WPQH experiments have been compiled on DIII-D, a machine-learning approach was taken to predict these ELMs. The complicated nature of this singular-event prediction task compelled us to utilize a wide-swath of diagnostics, including measurements of: turbulence (beam-emission spectroscopy [BES]), line-averaged density (CO$_2$ interferometry), local temperature gradients (electron-cyclotron emission [ECE]), $D_\alpha$ light emission, and actuator information from the neutral beams, ECH, and gas. 

A deep survival machine learning model to predict sporadic breakthrough ELMs was trained offline with the Auton Survival Model framework \cite{nagpal_auton-survival_nodate} using each of the aforementioned diagnostic measurements as input. Detailed documentation on the model, training, and database study is provided in an upcoming paper by \citet{farre_interpreting}.  The trained model was converted to 4 \texttt{C} functions using \texttt{keras2c}. As documented by \citet{conlin_keras2c_2021}, the outputted \texttt{C} functions are designed to be real-time compatible. Additional care was taken to ensure inputs provided to the \texttt{keras2c}-generated functions and internal math functions are bounded to avoid overflow. Based on the range of timescales of the prediction and control tasks, spanning the model's median prediction time of $\sim$\SI{100}{ms} and the model's execution time of $\sim$200$\mu$s, a short CPU cycle time of \SI{2}{ms} was selected. The output from the model block is (1-survivability), which represents the likelihood of ELM-free termination over a user-configurable time-horizon. The model block output is provided to the ELM-free PCS category's ELM avoidance algorithm, which conducts actuator decision-making.

For integrated testing, the model block also features a user-configurable debug waveform, which can be toggled to override the ML-model prediction to conduct integrated tests for inter-PCS algorithm communication and controller-side prediction handling.

\subsection{Alf\'en Eigenmode Controller}
An ML-based Alf\'en Eigenmode (AE) detection model was trained to classify different types of AEs based on ECE data\cite{jalalvand_alfven_2022,rothstein_initial_2024}. This observer model is an RCN-based ML model and was combined with basic proportional control (P-control) of $P_{NBI}$ to directly control the presence of AEs. This controller was tested and deployed in the ramp-up of a DIII-D shot where AEs are typically observed. 

The inputs for the AE detection model are purely ECE data. As shown in \citet{jalalvand_alfven_2022}, the performance of the model is generally stable to the frequency of data and so we receive data at the CPU cycle time on the order of \SI{1}{ms}, or \SI{1}{kHz}. The ECE data is received at \SI{1}{MHz} and is down-sampled without impacting the accuracy of the model. The output of the controller is $P_{NBI}$ and because NBI power directly affects the energetic particle distribution, this actuator will directly affect the AE amplitudes. The target CPU cycle time was approximately \SI{5}{ms} as that about the minimum timescale to change $P_{NBI}$ and was easily achieved by the RCN implementation. 

The model block contains the real-time RCN implementation which involved implementing a spare matrix multiplication algorithm necessary to deal with the highly sparse nature of the RCN design\cite{tanaka_recent_2019}. Unlike \texttt{keras2c}-generated functions, the weights for the RCN were loaded  as matrices to the PCS GUI and could therefore be changed during experiment to try out multiple models or retrain between experimental sessions. While other works have shown the possibility of retraining RCN type models intra-shot\cite{jalalvand_real-time_2021}, this has yet to be implemented but remains a possibility for future work. The model block will output AE "amplitudes" or the likelihood of that AE type currently being present in the plasma. The 5 types of AEs that are detected are: Beta-induced AE, Ellipicity-induced AEs, low-frequency modes, reversed shear AEs (RSAEs), and Toroidal AEs (TAEs). Based on the offline frequency of occurrences and accuracy of the detector model, the model only passes the signal for TAEs and RSAEs to the controller block. 

The controller block takes in the TAE and RSAE amplitudes and the user will decide which amplitude to use for feedback control as well as apply a low pass filter to smooth the RCN model outputs. The change in NBI power, $\Delta P_{NBI}$, is calculated by $\Delta P_{NBI}=K_P\times (Target_{AE}-Reference_{AE})$ where $K_P$ is the user-tuned proportional gain, $Target_{AE}$ is the target AE amplitude, and $Reference_{AE}$ is the RCN output amplitude. $\Delta P_{NBI}$ is then passed along to the NBI controller algorithm to adjust NBI duty cycles to achieve the requested NBI power target. There are limits to minimum and maximum $P_{NBI}$ that constrain the controller as well as limits to the maximum rate of change of $P_{NBI}$ to further constrain the controller as needed. 

\subsection{MPC Profile Control}\label{sec:MPC}

An AI-based Model Predictive Controller (MPC) is implemented for plasma profile prediction and control. Data-driven plasma profile predictors have shown great promise in predicting the evolution of a discharge\cite{abbate_data-driven_2021,char_full_2024}. These plasma models may be used for control through linearizing in the latent space and applying linear MPC\cite{otto_linearly_2019,watter_embed_2015}. This control algorithm is implemented for multi-input multi-actuator profile control, and has been tested experimentally in DIII-D producing accurate rotation and density 33-point full profile control. A more detailed publication is found in \citet{shousha_realtime}

The inputs for the profile controller model were the profiles provided by \texttt{RTCAKENN}, $I_p$, $B_T$, actuation hardware limits and the target profiles. The output of the controller block is NBI power and torque, ECH power and D2 gas injection voltage. The latent-space linearization encoder runs in 4ms, followed by the quadratic program (QP) solver for MPC taking up to 10ms, enabling a cycle time of 20ms. 

The model block consists of an MLP \texttt{keras2c}-generated function that maps the \texttt{RTCAKENN} point profiles into a linear latent space whose size depends on the model training, but is typically around 10. The same mapping is applied to the user-input target profiles, producing a measured latent space and a target latent space.

The controller block uses the measured latent space, target latent space, $I_p$, $B_t$, actuator hardware limits and QP solver parameters and runs a QP optimization to output the optimal NBI power, torque, ECH power and gas voltage to reach the profile targets.

\section{Conclusion}\label{sec:conclusion}
The \texttt{PACMAN} algorithm provides a testing environment for novel ML-based models and controllers that leverage the diverse set of DIII-D diagnostic data to perform advanced real-time control. The input block provides basic diagnostic pre-processing to ensure adequate information without excessive constraints so ML models still have significant flexibility in their own normalizations and additional pre-processing. The model block allows for multiple types of ML model implementation with the opportunity for new model types to be developed and deployed in the future. The controller block takes the outputs from the various ML models and uses appropriate standard or advanced control techniques to produce actuator commands. Finally, the output block handles conflicts in actuator requests and ensures the tokamak's safety by imposing strict actuator constraints. The clear separation of each block eases the user design process and allows for the possibility of each block to be parallelized in the future. 

The \texttt{PACMAN} framework has been demonstrated successfully in 5 listed experiments as well as an additional 4 not described in this manuscript: a divertor detachment controller \cite{chen_regulation}, an RL-based TM controller\cite{seo_avoiding_2024}, a finite-state machine profile controller\cite{abbate_data-driven_2021}, and a previous RL $\beta_N$ controller\cite{char_offline_2023}. The listed diverse experiments cover multiple types of model implementations utilizing \texttt{keras2c} for implementation of ML-based survival models, non-linear encoders and RCN for event prediction models. There were different types of control as well including fully ML-based RL control, finite state machines, PID-based controllers, and advanced MPC controllers. The variability in model and controller types demonstrates \texttt{PACMAN}'s flexibility to handle different control types and objectives that utilize the many diagnostics and actuators DIII-D has at its disposal. 

In its current form, \texttt{PACMAN} is prepared for the inclusion of new models and controllers to achieve DIII-D future control milestones. The biggest future goal is to implement the multi-threading of the entire model and controller blocks to speed up the computation time of the full algorithm. Additionally, as new diagnostics and actuators come online, the inputs and output blocks will be adjusted to include those new control capabilities. Future work can also explore coupling GPU hardware or possibly FPGAs with \texttt{PACMAN} and the DIII-D PCS to further speed-up real-time inference speeds.

\section*{Credit Statement}
\begin{itemize}
    \item[] \textbf{A. Rothstein}: Lead Author. Design Lead. Manuscript Lead. TM Control Lead. AE Control Co-Lead.
    \item[] \textbf{H.J. Farre-Kaga}: Lead Author. Design Lead. Profile Control Lead
    \item[] \textbf{J. Butt}: ELM Predictor Lead. Design support. PCS Testing Lead.
    \item[] \textbf{R. Shousha}: \texttt{RTCAKENN} Lead. Design Support. Paper editing. 
    \item[] \textbf{K. Erickson}: PCS Programming Lead. 
    \item[] \textbf{T. Wakatsuki}: RL ITB Lead.
    \item[] \textbf{A. Jalalvand}: Design Support. AE Control Co-Lead. High-frequency diagnostics lead.
    \item[] \textbf{P. Steiner}: Design Support. Paper editing. 
    \item[] \textbf{S.K. Kim}: Design Support. 
    \item[] \textbf{E. Kolemen}: Design Support. Supervision. Funding Support.  
\end{itemize}

\section*{Acknowledgment}

This material is based upon work supported by the U.S. Department of Energy, Office of Science, Office of Fusion Energy Sciences, using the DIII-D National Fusion Facility, a DOE Office of Science user facility, under Award DE-FC02-04ER54698. Additionally, this material is supported by the National Science Foundation Graduate Research Fellowship under Grant No. DGE-2039656 and by the U.S. Department of Energy, under Awards DE-SC0015480 and DE-AC02-09CH11466. 

We would also like to thank David Eldon for insightful feedback on this manuscript. 

\section*{Disclaimer}

This report was prepared as an account of work sponsored by an agency of the United States Government. Neither the United States Government nor any agency thereof, nor any of their employees, makes any warranty, express or implied, or assumes any legal liability or responsibility for the accuracy, completeness, or usefulness of any information, apparatus, product, or process disclosed, or represents that its use would not infringe privately owned rights. Reference herein to any specific commercial product, process, or service by trade name, trademark, manufacturer, or otherwise does not necessarily constitute or imply its endorsement, recommendation, or favoring by the United States Government or any agency thereof. The views and opinions of authors expressed herein do not necessarily state or reflect those of the United States Government or any agency thereof.

\section*{References}
\bibliographystyle{plainnat}
\bibliography{PACMAN.bib}

\section{Appendix}\label{sec:appendix}
\subsection{ML Controller Planning}\label{sec:planning}

Before exploring the controller examples that have been deployed in experiment using the \texttt{PACMAN} framework, we will provide a summary of important points within the \texttt{PACMAN} framework that are relevant for a real-time experimentalist designing and implementing a ML model for real-time plasma control. This is also relevant for exploring other non-CPU based implementations of ML models as similar design considerations will come into play. 

The real-time implementation design process must begin before any ML model is trained in order to achieve the best success. Without prior consideration of these design decisions, the experimentalist will likely need to retrain their model from scratch or make significant changes to their control scheme. The experimentalist should have, at minimum, the following considerations: 
\begin{itemize}
    \item What is the control objective? What timescale does this control objective evolve on? Is this compatible with approximate timescales for real-time ML models? 
    \item What actuators will be used to reach this control objective? What timescale do my actuator need to change to achieve this target? 
    \item What inputs will the ML model use? What is the sim-to-real gap between the real-time data and the offline training data? How can that sim-to-real gap be minimized? How quickly does the model need to run to achieve my control objectives?
    \item What type of control scheme will be used with the ML model? Does this controller know about actuator limits and actuator failures?  
\end{itemize}

Utilizing the \texttt{PACMAN} framework provides some basic answers to some of these questions in order to guide a novice interested in deploying a real-time control algorithm. The main forced answers come into play by imposing limits to timescales, forcing all model evaluation, control actuation, and control objective timescales to be on the order of milliseconds or longer for control and objective timescales. While this is insufficient for all plasma control applications, the \texttt{PACMAN} framework covers the vast majority of relevant timescales and enables ease of deployment for a significant number of ML models and controllers.

\end{document}